\documentclass{mn2e}
\usepackage{color}
\usepackage{graphicx}
\usepackage{dcolumn}
\usepackage{bm}

\begin{document}

\title[The $M_{\rm bh}$-$L$ relation]
{The Black Hole Mass - Spheroid Luminosity relation}

\author[Alister W.\ Graham]
{Alister W.\ Graham$^{1}$\thanks{AGraham@astro.swin.edu.au}\\
$^1$Centre for Astrophysics and Supercomputing, Swinburne University
of Technology, Hawthorn, Victoria 3122, Australia
}

\date{Received 2008 Jan 01; Accepted 2008 December 31}
\pubyear{2006} \volume{000}
\pagerange{\pageref{firstpage}--\pageref{lastpage}}

\input{psfig}

\maketitle
\label{firstpage}

\begin{abstract}
  
The differing $M_{\rm bh}$-$L$ relations presented in McLure \& Dunlop,
Marconi \& Hunt and Erwin et al.\ have been investigated. 
A number of issues have been identified and addressed in each
of these studies, including but not limited to: 
the removal of a dependency on the Hubble constant; 
a correction for dust attenuation in the bulges of disc galaxies; 
the identification of lenticular galaxies previously treated as elliptical
galaxies; and application of the same ($Y$$\mid$$X$) regression analysis. 
These adjustments result in relations which now predict similar 
black hole masses.
The optimal $K$-band relation is
$\log(M_{\rm bh}/M_{\sun}) = -0.37(\pm0.04)[M_K +24] + 8.29(\pm0.08)$, 
with a total (not intrinsic) scatter in $\log M_{\rm bh}$ equal to 0.33 dex. 
This level of scatter is similar to the 
value of 0.34 dex from the $M_{\rm bh}$-$\sigma$ relation of Tremaine et al.\ and 
compares favourably with the value of 0.31 dex from the $M_{\rm bh}$-$n$
relation of Graham \& Driver.
Using different photometric data, consistent relations in the $B$- and
$R$-band are also provided, 
although we do note that the small ($N=13$) $R$-band sample 
used by Erwin et al.\ is found here to have a slope of $-0.30\pm0.06$ 
and a total scatter of 0.31 dex. 
Performing a symmetrical regression on the larger $K$-band sample 
gives a slope of $\sim -0.40$,
implying $M_{\rm bh} \propto L^{1.00}$. 
Implications for galaxy-black hole coevolution, in terms of dry
mergers, are briefly discussed, as are the predictions for
intermediate mass black holes. 
Finally, as previously noted by Tundo et al., a potential bias in the galaxy
sample used to define the $M_{\rm bh}$-$L$ relations is shown and a 
corrective formula provided.

\end{abstract}


\section{Introduction}

The mass of a spheroid's supermassive black hole (SMBH), 
denoted by $M_{\rm bh}$, is known to
correlate with several physical properties of the spheroid, by which we mean
either an elliptical galaxy or the bulge of a disc galaxy. 
At first it was thought that the velocity dispersion, $\sigma$, of the 
bulge was the primary driving mechanism 
(Ferrarese \& Merritt 2000; 
Gebhardt et al.\ 2000).  However, in the following year 
the radial concentration of stars in the spheroids
was observed to correlate just as well with the SMBH mass
and yielded a relation with the same small degree of scatter 
(Graham et al.\ 2001). 
While the luminosity of the stars initially appeared to provide 
a weaker relation, it is now known that this was predominantly due to 
rough estimates of the spheroid's luminosity.  Performing a S\'ersic 
(1963) $R^{1/n}$-bulge plus exponential disc
decomposition of the galaxy light, Marconi \& Hunt (2003) and Erwin 
et al.\ (2004) revealed that the total scatter about the near-infrared and 
optical $M_{\rm bh}$-$L$ relations was similarly small at $\sim$0.35 dex. 
%
Most recently, Graham \& Driver (2007a, their Section~6) 
have predicted that the central
stellar density, prior to core-depletion of the host spheroid, may also 
be intimately connected with the SMBH mass. 


These relationships are important for two main reasons.
First, they provide an easy means to predict SMBH masses in 
thousands of galaxies for which direct measurements of the SMBH mass is not
possible (e.g., Yu \& Tremaine 2002; Marconi et al.\ 2004; 
Shankar et al.\ 2004; Graham et al.\ 2007 and references therein).  
The other reason is that they are a clue to the driving 
physical processes at work in galaxies. 

However, as with the controversy over the slope of the $M_{\rm bh}$-$\sigma$
relation (Merritt \& Ferrarese 2001a; Tremaine et al.\ 2002; Novak et al.\
2006), the slope of the $M_{\rm bh}$-$L$ relation is not yet agreed upon.
While McLure \& Dunlop report a value of $-0.50\pm0.05$, the data
in Erwin et al.\ (2004) has a slope of $-0.25\pm0.05$. 
Moreover, the differing relations from different studies tend not to 
predict the same SMBH mass. 
For example, when
$L_K = 10^{10}L_{K,\sun}$ ($10^{11}L_{K,\sun}$), the 
$K$-band expression in 
McLure \& Dunlop (2004, their equation~1)
predicts SMBH masses which are three 
(two) times less than those predicted by the $K$-band $M_{\rm bh}$-$L$
relation in Marconi \& Hunt (2003). 

In this paper we re-investigate the past $M_{\rm bh}$-$L$ relations and 
address a number of ways in which they can be updated.  
In Section~2 we present an early data set used to construct an 
$M_{\rm bh}$-$L$ relation.  We use this data to illustrate how 
we define the regression analysis that we shall adopt in this paper. 
Section~3 starts with studies which 
avoided the bulge/disc separation issue by excluding disc galaxies.
while Section~4 explores those studies which used elliptical, lenticular
and spiral galaxies. 
We obtain new $B$-, $R$- and $K$-band relations which are consistent
with each other and suitable for predicting black hole masses in 
other galaxies.   
In Section~5 we discuss 
how intermediate mass black holes factor in, and 
briefly mention some of the implications of these
new relations for the coevolution of spheroids and SMBHs, 

Finally, in an appendix we provide a re-derivation of the
$M_{\rm bh}$-$L$ relation taking into allowance a potential
bias in the galaxy sample.

\section{Construction of the $M_{\rm bh}$-$L$ relation}

\subsection{Kormendy \& Gebhardt (2001)}

As noted by Kormendy \& Gebhardt (2001, see their section~5) 
SMBHs seem to be associated with the dynamically hot, 
spheroidal component of a galaxy.  Those authors therefore 
presented a $B$-band relation based on the bulge, rather 
than total, magnitudes from Faber et al.\ (1997). 
Using $M_{B,\sun}=5.47$ mag, their $M_{\rm bh}$-$L$ relation can be written as 
\begin{equation}
\log(M_{\rm bh}/M_{\sun}) = -0.43[M_B + 19.5] + 7.88. 
\end{equation}

Given our objective is to construct the optimal relation ($y=a+bx$) for
estimating the SMBH mass ($y$) from the magnitude ($x$) of a galaxy, we adopt
the method of regression analysis given in Tremaine et al.\
(2002)\footnote{This 
is not a symmetrical method of regression, but rather one which minimises
the scatter in the $y$-direction (see Novak et al.\ 2006 and Graham \& Driver
2007a).  The criticism in 
Tundo et al. (2007, their Section 2.2) of the Tremaine et al.\ (2002) 
method is therefore misplaced.} 
We allow for intrinsic scatter
(in the y-direction), which we denote by the term $\epsilon$, and also for
measurement errors
on the $N$ pairs of observables $x_i$ and $y_i$, which we denote $\delta x_i$
and $\delta y_i$.  Tremaine et al.'s (2002) modified version of the routine
FITEXY (Press et al.\ 1992, their Section~15.3) minimises the quantity 
\begin{equation}
\chi^2 = \sum_{i=1}^N \frac{( y_i- a - bx_i)^2}
    { {\delta y_i}^2 + b^2{\delta x_i}^2 + \epsilon^2 }.
\label{Eq_One}
\end{equation}
The intrinsic scatter $\epsilon$ is solved for by repeating the fit until
$\chi^2/(N-2)$ equals 1.  The uncertainty on $\epsilon$ is obtained when the
reduced chi-squared value, $\chi^2/(N-2)$, equals $1 \pm \sqrt{2/N}$.  To
achieve a minimisation in the $x$-direction, one simply replaces the
$\epsilon^2$ term in the denominator of equation~\ref{Eq_One} with
$b^2\epsilon^2$.  A symmetrical regression is therefore some kind of average
of these two regressions (Novak et al.\ 2006). 

Application of equation~\ref{Eq_One} to the data from Kormendy \& Gebhardt
yields the relation 
\begin{equation}
\log(M_{\rm bh}/M_{\sun}) = -0.38(\pm 0.06)[M_B + 19.5] + 8.00(\pm 0.09), 
\end{equation}
which has a total rms scatter of 0.56 dex in the $\log M_{\rm bh}$ direction 
and an intrinsic scatter (assuming a magnitude error of 0.3 mag) of 
$0.46^{+0.08}_{-0.06}$ dex. 

This level of scatter is unpleasantly high and resulted in the 
$M_{\rm bh}$-$L$ relation taking second place to the $M_{\rm bh}$-$\sigma$ 
(and $M_{\rm bh}$-$n$) 
relation.  However, it has since been realised/shown that this level of scatter 
was a consequence of a poor bulge/disc separation (Marconi \& Hunt 2003; 
Erwin et al.\ 2004) and the use of systems 
in which the SMBH's sphere of influence was not well resolved (e.g.\
Merritt \& Ferrarese 2001c; Ferrarese \& Ford 2005).
The following Section avoids the issue of the bulge/disc separation
by dealing with a study that used elliptical galaxies.  Section~\ref{SecDisc}
effectively tackles this issue by using studies in which a
S\'ersic-bulge\footnote{A modern review of the S\'ersic model can be found in
  Graham \& Driver (2005).} plus exponential-disc decomposition of the
galaxy's stellar light has been performed.

%

\section{Elliptical galaxies}

\subsection{McLure \& Dunlop (2002)} \label{SecMD}

Given that the mass of a SMBH is known to correlate with the
properties of the host spheroid, rather than the host galaxy, McLure \& 
Dunlop (2002) presented the $M_{\rm bh}$-$L$ relation after excluding 
(the bulk of the) disc galaxies\footnote{Bettoni et al.\ (2003) 
performed the same task, obtaining a consistent relation.}.
Their expression (for a predominantly elliptical galaxy 
sample) is
\begin{equation}
\log(M_{\rm bh}/M_{\sun}) = -0.50(\pm0.05)M_R - 2.91(\pm1.23).
\label{Eq_MD18}
\end{equation}
Here we re-derive this expression after implementing the following
alterations.
\begin{itemize}
\item 
  Measurements of black hole mass depend linearly on the distance to
  each galaxy.  Distances for 16 of the 18 galaxies used by McLure \&
  Dunlop have been obtained using surface brightness fluctuations (Tonry et al.\ 
  2001) and so their SMBH masses are independent of the Hubble constant.
  However, in converting the apparent magnitude of these 18 galaxies into
  absolute magnitudes, McLure \& Dunlop used 
  the galaxy redshift and a Hubble constant $H_0 = 50$ 
  km s$^{-1}$ Mpc$^{-1}$.  This was done because of their comparison with 
  a sample of AGN for which $H_0 = 50$ km s$^{-1}$ Mpc$^{-1}$ had been used. 
  A more consistent approach would involve the use of the $H_0$-independent 
  distances for the non-AGN galaxies to determine their absolute magnitudes, 
  and a Hubble constant of 73 km s$^{-1}$ Mpc$^{-1}$ (Blakeslee et al.\ 2002)  
  for the AGN sample.  Here we use these $H_0$-independent distances for 
  {\it both} the SMBH masses and the spheroid
  magnitudes of our (local non-AGN) galaxies. 

\item 
  Nine of the 18 $R$-band magnitudes which
  McLure \& Dunlop used were ($B-R_c = 1.57$)-adjusted $B$-band magnitudes from
  Faber et al.\ (1997) --- which themselves came from the ``Seven Samurai'' data
  (Faber et al.\ 1989) and/or the third reference galaxy catalogue (RC3, de
  Vaucouleurs et al.\ 1991).  The other nine $R$-band magnitudes are
  reported to be 
  ($V-R = 0.61$)-adjusted $V$-band magnitudes from Merritt \& Ferrarese
  (2001b).  
  Although this paper has no  $V$-band magnitudes, Merritt \& Ferrarese
  (2001a) has $V$-band magnitudes from Faber et al.\ (1989) for 4 of these
  9 galaxies.   We therefore use the Faber et al.\ (1989) $B$-band apparent magnitudes
  (and the related absolute magnitudes in Tremaine et al.\ 2002).
  The Faber et al.\ (1989) magnitudes are derived from photoelectric
  aperture growth curves.  While the magnitude quality indicators from that
  paper suggest an error of $<$ 0.15 or 0.30 mag, depending on the
  galaxy, we have adopted the upper value for all galaxies in our regression 
  analysis.  

\item 
  We have removed NGC~4564 and NGC~2778 which are not elliptical galaxies but S0
  galaxies\footnote{Lenticular galaxies have typical bulge-to-total luminosity
    ratios of 1/4 (e.g., Balcells, Graham \& Peletier 2004; Laurikainen, Salo \&
    Buta 2005).}  (see Graham \& Driver 2007b) whose total galaxy luminosities
  would have biased the previous relation.
  We have also excluded the peculiar elliptical galaxy IC~1459 due to
  uncertainty on its SMBH mass.  While the stellar dynamics of its core
  suggest a SMBH mass of $2.6 \times 10^9 M_{\sun}$, the gas dynamics reveal
  the mass could be as low as $3.5 \times 10^8 M_{\sun}$ (Cappellari et al.\ 
  2002).  We have treated NGC~221 as an S0 galaxy according to the $B/T$ 
  flux ratio in Graham (2002). 
  
\item 
  We have included NGC~1399 and NGC~5845 for which accurate SMBH masses
  have since become available, giving a total of 17 galaxies.

\item
  Due to our desire to obtain a relation for predicting accurate
  SMBH masses using the magnitudes of other galaxies, we perform a
  non-symmetrical regression which results in the smallest degree of scatter
  in the $\log M_{\rm bh}$ direction.  That is, we apply equation~\ref{Eq_One}. 

\end{itemize}

\begin{table}
\centering
\begin{minipage}{84mm}
\caption{
Revised (see Section~\ref{SecMD}) 
sample of elliptical galaxies from McLure \& Dunlop (2002). 
Distances are taken from Tonry et al.\ (2001, their table 1), except for 
NGC~6251 ($v_{\rm CMB}$=7382 km s$^{-1}$, Wegner et al.\ 2003) and
NGC~7052 ($v_{\rm CMB}$=4411 km s$^{-1}$, Wegner et al.\ 2003).
These two galaxies are not listed in Tonry et al.\ (2001) and 
a Hubble constant of $H_0 = 73$ km s$^{-1}$ Mpc$^{-1}$ (Blakeslee 
et al.\ 2002; Spergel et al.\ 2006) has been used. 
Unless noted otherwise,
the apparent $B$-band magnitudes, $m_B$, are the $B_T^0$ 
magnitudes from Faber et al.\ (1989, 1997).  Unless noted otherwise, the
absolute $B$-band magnitudes, $M_B$, have come from Tremaine et al.\ (2002)
with modifications for NGC~6251 and NGC~7052 such that we adjusted the
Tremaine et al.\ values ($-21.81$ and $-21.31$ mag) to our adopted distance.
The SMBH masses are also from the compilation in Tremaine et al.\ (2002), except for
NGC~821 (Richstone et al.\ 2007), 
NGC~3379 (Gebhardt et al.\ 2000; see also Shapiro et al.\ 2006), and 
NGC~4486 (Macchetto et al.\ 1997). 
Our sample includes two additional galaxies not used in Tremaine et al. 
The SMBH mass for NGC~1399 is from Houghton et al.\ (2006) and 
the mass for NGC~4374 is from Maciejewski \& Binney (2001, with
updated errors taken from Kormendy \& Gebhardt 2001).
\label{TabMD}
}
\begin{tabular}{@{}lcrcc@{}}
\hline
Galaxy       & Dist.            &  $m_B$ &  $M_B$  &  $M_{\rm bh}$       \\
             & [Mpc]            & [mag]  & [mag]   &  [$10^8 M_{\sun}$]  \\
NGC 221      & 0.81             & (9.28)\footnote{Reduced using a $B/T$ ratio
  of 0.62 (Graham 2002).}  & ($-14.50$)\footnote{$R$-band bulge 
magnitude from Graham (2002) with a $B-R=1.84$ mag adjustment (Lugger et al.\
1992.}  &  0.025$^{+0.005}_{-0.005}$ \\ 
NGC 821\footnote{SMBH sphere of influence not resolved.} 
             & 24.1             & 11.57  &  $-20.41$ &  0.85$^{+0.35}_{-0.35}$ \\
NGC 1399     & 20.0             & 10.55  & ($-20.96$)\footnote{Derived from the
  apparent magnitude.}  &  12$^{+5}_{-6}$  \\
NGC 3377\footnote{SMBH sphere of influence not resolved.}  
             & 11.2             & 11.13  &  $-19.05$ &  1.00$^{+0.9}_{-0.1}$  \\
NGC 3379     & 10.6             & 10.43  &  $-19.94$  &  1.35$^{+0.73}_{-0.73}$  \\
NGC 3608     & 22.9             & 11.68  &  $-19.86$ &  1.90$^{+1.0}_{-0.6}$  \\
NGC 4261     & 31.6             & 11.32  &  $-21.09$ &  5.20$^{+1.0}_{-1.1}$  \\
NGC 4291     & 26.2             & 12.42  &  $-19.63$ &  3.10$^{+0.8}_{-2.3}$  \\
NGC 4374     & 18.4             & 10.13  & ($-21.19$)\footnote{Derived from the 
  apparent magnitude.}  &  4.64$^{+3.46}_{-1.83}$  \\
NGC 4473     & 15.7             & 11.21  &  $-19.89$ &  1.10$^{+0.40}_{-0.79}$  \\
NGC 4486     & 16.1             &  9.52  &  $-21.53$ &  34.3$^{+9.7}_{-9.7}$  \\
NGC 4649     & 16.8             &  9.77  &  $-21.30$ &  20.0$^{+4.0}_{-6.0}$  \\
NGC 4697     & 11.7             & 10.03  &  $-20.24$ &  1.70$^{+0.2}_{-0.1}$  \\
NGC 4742     & 15.5             & 12.03  &  $-18.94$ & (0.14$^{+0.04}_{-0.05}$)\footnote{This 
            SMBH mass is based on M.E.\ Kaiser (2001, in prep.).} \\ 
NGC 5845     & 25.9             & 13.35  &  $-18.72$ &  2.40$^{+0.4}_{-1.4}$  \\
NGC 6251     & 101$h^{-1}_{73}$ & (13.64)\footnote{RC3 $B (m_B^0)$ value. (No
  value in Faber et al.\ 1989.)}  &  $-21.99$ &  5.80$^{+1.8}_{-2.0}$  \\
NGC 7052     & 60$h^{-1}_{73}$  & (12.73)\footnote{RC3 $B (m_B^0)$ value. (No
  value in Faber et al.\ 1989.)}  &  $-21.36$ &  3.40$^{+2.4}_{-1.3}$  \\
\hline
\end{tabular}
\end{minipage}
\end{table}

\begin{figure*}
\includegraphics[angle=270,scale=0.69]{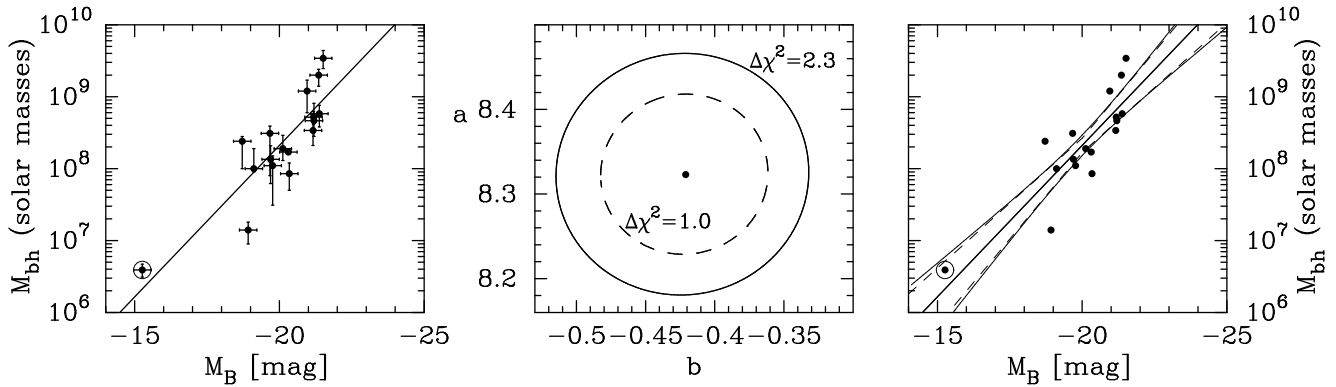}
\caption{
Correlation between an elliptical galaxy's supermassive black hole mass and
the apparent $B$-band magnitudes listed in Table~\ref{TabMD}. 
The regression line shown in the left panel was obtained using
equation~\ref{Eq_MD_M32}.
%
%
%
NGC~221 (M32) is plotted with a circle around it. 
The middle panel shows the $\Delta \chi^2=1.0$ and 2.3 boundaries
around the optimal intercept, $a=8.32$, and slope, $b=-0.42$.  The
projection of the $\Delta \chi^2=1.0$ ellipse onto the vertical and
horizontal axis gives the 1-$\sigma$ uncertainties $\delta a$ and
$\delta b$, respectively.  The $\Delta \chi^2=2.3$ ellipse denotes the
1-$\sigma$ {\it two-dimensional} confidence region.
This latter quantity has been mapped into the right panel, and is traced by the two
solid curves.  The dashed lines in this panel are the (more commonly
used) approximations obtained using ($a\pm\delta a$) and ($b\pm\delta
b$).  The two confidence regions agree well, although the region traced
by the dashed lines is, as expected, smaller.
}
\label{fig-MD17}
\end{figure*}

Table~\ref{TabMD} provides distances, updated SMBH masses 
(based on these distances) and absolute magnitudes (again based on
these distances) for this set of galaxies. 
%
%
Using the distances and apparent bulge magnitudes from Table~\ref{TabMD}, and applying 
equation~\ref{Eq_One}, 
we obtain the relation 
\begin{equation}
\log(M_{\rm bh}/M_{\sun}) = -0.42(\pm0.06)[M_B +20] +8.32(\pm0.10),
\label{Eq_MD_M32}
\end{equation}
with an absolute scatter in $\log M_{\rm bh}$ of 0.36 dex. 
This relation can be seen in Figure~\ref{fig-MD17}.
%
%
Using the absolute magnitudes given in Table~\ref{TabMD}, 
rather than the apparent magnitudes, one obtains 
\begin{equation}
\log(M_{\rm bh}/M_{\sun}) = (-0.36\pm0.06)[M_B +20] +(8.33\pm0.10). 
\label{Eq_Abs}
\end{equation}

These equations have an order of magnitude less uncertainty on the 
intercept term than the uncertainty given in equation~\ref{Eq_MD18}. 
They also have a shallower slope.

\subsubsection{Independence of $H_0$} \label{SecH0}

As can be seen in Table~\ref{TabMD}, 
two of the galaxies used to construct the above $M_{\rm bh}$-$L$ relation 
have magnitudes and black hole masses that depend on the Hubble constant.  We
have explored whether or not equation~\ref{Eq_MD_M32} and \ref{Eq_Abs}
would be significantly different if
(i) these two galaxies were excluded
and (ii) if we had used a Hubble constant of 50 
or 100, 
rather than 73
km s$^{-1}$ Mpc$^{-1}$ for these two galaxies (NGC~6251 and NGC~7052). 
In all cases the slope and intercept varied by no more than 0.02 and 0.03,
respectively.   What this means is that the $M_{\rm bh}$-$L$ relations 
given above, and in the rest of this Section, are effectively independent
of the Hubble constant.

\subsubsection{An $R$-band $M_{\rm bh}$-$L$ relation}

As already noted, the $R$-band relation from McLure \& Dunlop (2002) was based
on the Faber et al.\ (1989) $B$-band magnitudes and an $B-R_c$
colour\footnote{From here on the subscript $c$ shall be dropped from the 
    term $R_c$.} equal to 1.57 mag. 
Adopting this same colour term, except for NGC~221 which 
has a $B-R$ colour of 1.84 (Lugger et al.\ 1992), and using the absolute 
magnitudes given in Table~\ref{TabMD}, one obtains the relation 
%
%
\begin{equation}
\log(M_{\rm bh}/M_{\sun}) = -0.38(\pm0.06)[M_R +21] +8.11(\pm0.11). 
\label{Eq_MD17R}
\end{equation}

For a dynamically hot spheroid with $M_R = -21$ mag, McLure \& Dunlop's
relation (equation~\ref{Eq_MD18}) 
gives values of $\log M_{\rm bh}$ which are 0.52 dex less massive,
i.e.\ roughly a factor of three less massive, 
and considerably offset from our refined zero-point at 
$M_R = -21$ mag, given by $8.11\pm0.11$.   
However it should be noted that 
most of this offset (at $M_R = -21$ mag) 
is simply due to the prior use of $H_0 = 50$ km s$^{-1}$ Mpc$^{-1}$ 
when deriving the absolute magnitudes used in equation~\ref{Eq_MD18}. 
If a value of 73 km s$^{-1}$ Mpc$^{-1}$ is used there, then the masses
are now only 0.11 dex smaller, 
which is consistent with the 0.11 dex uncertainty on the new zero-point. 


\subsubsection{A $K$-band $M_{\rm bh}$-$L$ relation}

We also update the $K$-band relation presented
in McLure \& Dunlop (2004, their equation~1), which was derived there 
using an $R-K$ colour of 2.7 mag applied to their $R$-band relation 
from McLure \& Dunlop (2002). 
When McLure \& Dunlop (2004) converted McLure \& Dunlop's (2002) 
$M_{\rm bh}$-$L$ relation, they also 
applied a Hubble conversion of $\log(50/70)$ to the 
black hole masses and $5\log(50/70)$ to the magnitudes. 
This is not applied here because equation~\ref{Eq_MD17R} is, for 
practical purposes, independent of the Hubble constant (Section~\ref{SecH0}).
Adopting an $R-K$ colour of 2.6 mag (Buzzoni 2005), 
equation~\ref{Eq_MD17R} becomes 
\begin{equation}
\log(M_{\rm bh}/M_{\sun}) = -0.38(\pm0.06)[M_K +24] + 8.26(\pm0.11).
\label{Eq_MD17K}
\end{equation}
Using here an absolute $K$-band magnitude for the Sun 
of $M_{K,\sun}=3.28$, this can alternatively be expressed as 
\begin{equation}
\log(M_{\rm bh}/M_{\sun}) = 0.95(\pm0.15)\log \frac{L_{K,sph}}
 {10^{10.91} L_{K,\sun}}  + 8.26(\pm0.11), 
\end{equation}
where $L_{K,sph}/L_{K,\sun}$ is the $K$-band luminosity of the spheroid 
component of the galaxy (i.e., the bulge or the elliptical galaxy itself)
in solar units.

%

\section{Including the disc galaxies} \label{SecDisc}

More recent studies have included both elliptical galaxies and 
the bulges of disc galaxies. 
In this section we explore the $M_{\rm bh}$-$L$ expressions 
obtained with both sets of objects.

\subsection{Erwin, Graham \& Caon (2002)} \label{SecEGC}

In 2002 Erwin, Graham \& Caon presented a relation between black hole mass and
host spheroid magnitude for a sample of 13 galaxies (8 elliptical and 5 disc
galaxies).  The study was presented at the Carnegie Observatories Astrophysics
conference ``Coevolution of Black Holes and Galaxies'' and posted to astro-ph
that same year.
For the disc galaxies, the bulge magnitudes were obtained from an
$R^{1/n}$-bulge plus exponential-disc decomposition (Graham et al.\ 2001).
This was the first study, albeit small in number, to show that the inclusion
of bulge galaxies resulted in a total scatter of 0.35 dex in $\log M_{\rm bh}$,
considerably less than past reports of $\sim$0.6 dex and comparable to the
scatter in the $M_{\rm bh}$-$\sigma$ and $M_{\rm bh}$-$n$ relations.

\begin{figure}
\includegraphics[angle=270,scale=0.39]{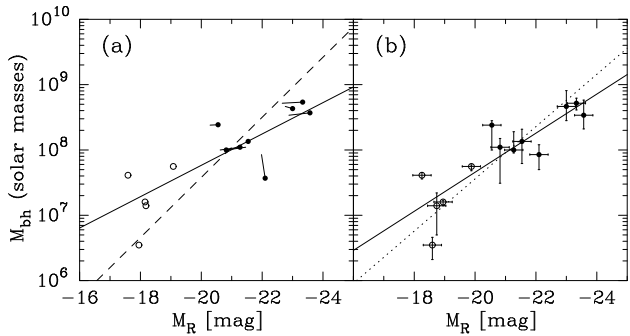}
\caption{
The long solid line shows the $M_{\rm bh}$-$L$ 
relation (equation~\ref{EqErwin} and \ref{Eq_Ali}) using the 8 elliptical 
galaxies (filled circles) plus the bulges of 5 lenticular/spiral 
galaxies (open circles) given in Erwin et al.\ (2004).  The short
lines emanating from the elliptical galaxy data points in panel a) 
show the location of these galaxies as used in Section~\ref{SecMD} to obtain 
equation~\ref{Eq_MD17R} (shown by the long dashed line).
Panel a) shows the data points as seen in Erwin et al.\ (2004; their Figure~1.3). 
Panel b) shows the data points as given in Table~\ref{Tab_EGC},
which uses slightly updated SMBH masses and, most importantly, dust-corrected
bulge magnitudes (see Section~\ref{SecDust}).  The dotted line
has a slope of $-0.4$. 
}
\label{FigEGC}
\end{figure}

Using the data points shown in Figure~1.4 of Erwin et al. (2004), along with a
0.3 mag uncertainty on the magnitudes, we have performed a regression of $\log
M_{\rm bh}$ against these $R$-band magnitudes (i.e., applied
equation~\ref{Eq_One}).  Doing so gives
%
%
\begin{equation}
\log(M_{\rm bh}/M_{\sun}) = -0.24\pm0.05[M_R +21] +8.01\pm0.11, 
\label{EqErwin}
\end{equation}
with a total scatter of 0.35 dex.  This is shown in Figure~\ref{FigEGC}a.

This relation has a noticeably shallower slope than the value 
in equation~\ref{Eq_MD17R} of $-0.38(\pm0.06)$
obtained using our refined elliptical galaxy sample from McLure \& Dunlop
(2002).
When $M_R = -24$ mag, equation~\ref{EqErwin} gives masses of $\sim5\times 10^8
M_{\sun}$ --- three times smaller than that obtained using
equation~\ref{Eq_MD17R}. 
When $M_R = -27$ mag, equation~\ref{EqErwin} gives masses of 
$\sim3\times 10^9 M_{\sun}$. 
In this regard, equation~\ref{EqErwin} does not conflict with the 
$M_{\rm bh}$-$\sigma$ relation.  That is, compared to the 
$M_{\rm bh}$-$\sigma$ relation, equation~\ref{EqErwin}
does not predict significantly larger SMBH masses at the high mass end. 
However, even if the S\'ersic bulge (and elliptical galaxy) magnitudes used in
Erwin et al.\ are reliable, the low number of points (only 13) may make
equation~\ref{EqErwin} prone to statistical fluctuations in the selected
sample.

In Figure~\ref{FigEGC} we show the data points used by Erwin et al.\ (2004).
Also shown, for the elliptical galaxies, is their location using the values
adopted in the previous Section where a slope of $-0.38$ was obtained.  The
only data point which has shifted significantly is NGC~821.  The reason is
because the previous section used the SMBH mass from Richstone et al.\ (2007,
$8.5\times 10^7 M_{\sun}$), while Erwin et al.\ used the (at the time
unpublished) value of $3.7\times 10^7 M_{\sun}$ (Tremaine et al.\ 2002).

\subsubsection{Dust} \label{SecDust}

While, in general, elliptical galaxies and the bulges
of disc galaxies are considered not to have dust, this does
not mean that the dust in the discs of disc galaxies does not
influence the emergent flux from the stars in the bulge. 
To illustrate this, 
consider an infinitely thin and optically opaque dust sheet running through
the disc of a galaxy.  Obviously, near edge-on orientations aside, one
will only see the bulge (and disc) stars on the near-side of the disc.  
That is, one will
only see half of the optical flux from the bulge, it will therefore be
observed to be 0.75 mag fainter than it actually is.  In reality, dust discs
have a certain thickness, relative to the vertical scale-height of the
stars in the disc.  This results in a dust correction that depends on 
the inclination of the disc relative to our line of sight. 

Here we modify the 
$B$-band inclination-attenuation correction for bulge magnitudes
given in Driver et al.\ (2007, their equation~3). 
Specifically, we use the expression 
\begin{equation}
(M_{\rm obs}-M_{\rm dust-free})_{R, {\rm bulge}} = \\
 0.6 \times \left[ 0.84 + 2.16\left( 1-\frac{b}{a} \right)^{2.48} \right], 
\label{Eq_dust}
\end{equation}
in which the 0.84 term provides the face-on attenuation-correction 
to the observed magnitude of the bulge, $M_{\rm obs}$, 
while the latter part of the expression 
provides the inclination-dependent component of the correction.  
Equation~\ref{Eq_dust} only differs from the $B$-band relation 
due to our (simplistic) use of a 0.6 multiplier. 
This multiplicative factor stems from our knowledge of the disc extinction/attenuation 
in the $B$- and $R$-bands (e.g.\ Tully \& Verheijen 1997; Tully et al.\
1998). 

The five bulge magnitudes listed in Table~\ref{Tab_EGC} are corrected for
dust using equation~\ref{Eq_dust}.  The values plotted in Erwin et al.\ 
(2004), and the left panel of Figure~\ref{FigEGC} are some 0.55 to 0.80 mag
fainter. 
Re-performing the regression analysis, with the dust-corrected magnitudes, 
and using the Richstone et al.\ (2007) SMBH mass for NGC~821, 
we obtain
\begin{equation}
\log(M_{\rm bh}/M_{\sun}) = -0.30(\pm0.06)[M_R +21] + 7.96(\pm0.10), 
\label{Eq_Ali}
\end{equation}
with a scatter of 0.31 dex in $\log M_{\rm bh}$.  This is shown in
Figure~\ref{FigEGC}b. 

We have repeated this regression 
after the jackknife removal of NGC~2787, a galaxy with an inner
disc twice as luminous as its bulge 
(Erwin et al.\ 2003), and the faintest object in our sample. 
Doing so results in a slope and intercept of $-0.35\pm0.05$ and 
$7.90\pm0.09$, respectively. 
and a total scatter of 0.29 dex. 
%

To better gauge the uncertainty on the slope in equation~\ref{Eq_Ali},
we have also used a bootstrap
sampling of 13 data points (i.e.\ sampling with replacement from the original
sample) and 1000 such Monte Carlo samples.  
%
%
The median $\pm2\sigma$ of the resultant distribution of 1000 slopes was
$-0.31\pm0.11$.  For the intercept we obtained $7.96^{+0.16}_{-0.17}$.

\begin{table}
\caption{
Revised sample of elliptical and disc galaxies from Erwin et al.\ (2004). 
A galaxy Type `E' denotes an elliptical galaxy 
while a Type `S' denotes either an S0 galaxy or a spiral galaxy. 
Distances are taken from Tonry et al.\ (2001, their table 1), except for 
NGC~7052 ($v_{\rm CMB}$=4411 km s$^{-1}$, Wegner et al.\ 2003) which 
is not listed in Tonry et al.
A Hubble constant of $H_0 = 73$ km s$^{-1}$ Mpc$^{-1}$ 
has been used for this galaxy. 
The SMBH masses are from the compilation in Tremaine et al.\ (2002), 
except for NGC~821 (Richstone et al.\ 2007) and 
NGC~3379 (Gebhardt et al.\ 2000).
Our sample includes one additional galaxy not used in Tremaine et al.
The SMBH mass for NGC~4374 is from Maciejewski \& Binney (2001, with
updated errors taken from Kormendy \& Gebhardt 2001).
Disc inclinations, i.e.\ $b/a$ axis ratios of the outer isophotes,
have come from: 
NGC~2778 (Rix et al.\ 1999);  
NGC~2787 (Erwin et al.\ 2003); 
NGC~3384 (Faber et al.\ 1997);  
NGC~4564 (Faber et al.\ 1997, see also Graham \& Driver 2007a); 
and NGC~7457 (Chapelon et al.\ 1999).  The magnitudes are from Erwin et al.\
(2004),
but corrected for dust attenuation using equation~\ref{Eq_dust}. 
\label{Tab_EGC}
}
\begin{tabular}{@{}llrccl@{}}
\hline
Galaxy   & Type & Dist.  & $b/a$ & $M_R$  &  $M_{\rm bh}$      \\
         &      & [Mpc]  &       & [mag]  & [$10^8 M_{\sun}$]  \\
NGC 0821 &  E   &  24.1  &  ...  & -22.10 &  0.85$^{+0.35}_{-0.35}$ \\
NGC 3377 &  E   &  11.2  &  ...  & -21.27 &  1.00$^{+0.9}_{-0.1}$   \\
NGC 3379 &  E   &  10.6  &  ...  & -21.54 &  1.35$^{+0.73}_{-0.73}$ \\
NGC 4261 &  E   &  31.6  &  ...  & -23.33 &  5.20$^{+1.0}_{-1.1}$   \\
NGC 4374 &  E   &  18.4  &  ...  & -23.00 &  4.64$^{+3.46}_{-1.83}$ \\
NGC 4473 &  E   &  15.7  &  ...  & -20.82 &  1.10$^{+0.40}_{-0.79}$ \\
NGC 5845 &  E   &  25.9  &  ...  & -20.55 &  2.40$^{+0.4}_{-1.4}$   \\
NGC 7052 &  E   & 60$h_{73}^{-1}$ &...& -23.57 & 3.40$^{+2.4}_{-1.3}$ \\
NGC 2778 &  S   &  22.9  &  0.72 & -18.74 &  0.14$^{+0.08}_{-0.09}$  \\
NGC 2787 &  S   &   7.5  &  0.57 & -18.25 &  0.41$^{+0.04}_{-0.05}$  \\
NGC 3384 &  S   &  11.6  &  0.45 & -18.95 &  0.16$^{+0.01}_{-0.02}$  \\
NGC 4564 &  S   &  15.0  &  0.45 & -19.88 &  0.56$^{+0.03}_{-0.08}$  \\
NGC 7457 &  S   &  13.2  &  0.59 & -18.60 &  0.035$^{+0.011}_{-0.014}$ \\
\hline
\end{tabular}
\end{table}

\subsection{Marconi \& Hunt (2003)} \label{Sec_MH}

\subsubsection{A $K$-band $M_{\rm bh}$-$L$ relation}\label{SecMH_K}

Using images from {\it 2MASS}\footnote{Two micron all sky survey: Jarrett et
  al.\ (2000).}, Marconi \& Hunt (2003) obtained $K$-band magnitudes for 27
galaxies which had direct and reliable SMBH mass measurements.  Recognising the
need for bulge magnitudes rather than galaxy magnitudes, they also performed
an $R^{1/n}$-bulge plus exponential-disc decomposition of the disc galaxies in
their sample.  With a sample size twice that used in Erwin et al.\ (2004), 
they confirmed Erwin et al.'s claim for a scatter of around 0.35 dex
in the $M_{\rm bh}$-$L$ relation for elliptical galaxies and the bulges
of disc galaxies. 
In what follows we describe the manner and reason for why we 
have tweaked the masses and magnitudes for some of their galaxy sample. 

\begin{itemize}
  
\item Due to the shallow nature of the {\it 2MASS} images (only 8 seconds) the
presence of a disc in NGC~2778\footnote{Because we used Marconi \& Hunt's
  preferred 27 ``Group 1'' galaxies (see their Table~1), NGC~2778 does not
  actually factor into the analysis here because it is tabulated as a
  ``Group 2'' galaxy.}  and NGC~4564 were missed.  NGC~221 
(M32) was also
treated as a pure elliptical galaxy.  From the $R$-band bulge/disc
decomposition of these three galaxies in Graham \& Driver (2007a) and Graham
(2002), the bulge-to-total flux ratios are 0.21, 0.24 and 0.62 respectively.
We have used these ratios to obtain the bulge magnitudes for these
galaxies. 
In passing we note that lenticular galaxies typically have $B/T$ ratios
of $\sim0.25\pm0.10$ (Balcells et al.\ 2004; Laurikainen et al.\ 2005), the 
higher $B/T$ value for M32 is to be expected if this is a partially disc-stripped
lenticular galaxy (see Bekki et al.\ 2001; Graham 2002). 

\item
We have excluded IC~1459 due to the order of magnitude uncertainty
on its black hole mass (Cappellari et al.\ 2002), reducing our 
sample size from 27 to 26 galaxies (see Table~\ref{TabMaH}). 

\item
We have updated the SMBH mass and its associated uncertainty for 
NGC~5252 using the now published result in Capetti et al.\ (2005) 
together with a distance of 94.4 Mpc (slightly different to 
the value of 96.8 Mpc in Marconi \& Hunt, 
and obtained using a recession velocity 
of 6888 km s$^{-1}$ and $H_0 = 73$ km s$^{-1}$ Mpc$^{-1}$) to give 
$M_{\rm bh} = 0.97^{+1.49}_{-0.46} \times 10^9 M_{\sun}$.  While this 
distance and mass is only 2.5 per cent smaller than that given in 
Marconi \& Hunt, the uncertainty on the mass is 2-3 times larger. 

\item
We have also slightly modified the SMBH mass for Cygnus~A 
by using the value $M_{\rm bh} = (2.6\pm0.7) \times 10^9 M_{\sun}$.
This was obtained from the mass in Tadhunter et al.\ (2003) after using 
a redshift of 0.056 and adopting $H_0 = 73$ km s$^{-1}$ Mpc$^{-1}$
together with $\Omega_m=0.3$ and $\Omega_{\Lambda}=0.7$. 
This gave a luminosity distance of 240 Mpc 
and an angular distance of 215 Mpc (c.f.\ 207 Mpc in Tadhunter et al.). 
This mass is still consistent with the value of $(2.9\pm0.7) \times 10^9
M_{\sun}$ used in Marconi \& Hunt (2003). 

\item
In addition to NGC~5252 and Cygnus~A, Marconi \& Hunt's 
sample includes a further four galaxies not listed in Tremaine et al.\ (2002).
The SMBH mass we 
have used for NGC~3031 (M81) is from Devereux et al.\ (2003), 
the mass for NGC~5128 (Cen~A) has come from Marconi et al.\ (2001) and 
the mass for NGC~4594 is from Kormendy et al.\ (1988). 
The mass for NGC~4374 (M84) has been taken from Maciejewski \& Binney (2001, with 
updated errors taken from Kormendy \& Gebhardt 2001). 
These masses are the same as used by Marconi \& Hunt, except for NGC~4374. 

\item
For the remaining 20 galaxies we have adopted the SMBH masses given in Tremaine
et al.\ (2002), with only the following three exceptions: 
NGC~3379 (Gebhardt et al.\ 2000; see also Shapiro et al.\ 2006), 
NGC~4486 (M87, Macchetto et al.\ 1997) and 
NGC~3115 (Emsellem, Dejonghe, \& Bacon 1999).
Our mass for NGC~3379 is 35 per cent larger than that used by Marconi \& Hunt,
while the difference in mass for the other two galaxies is only $\sim$1 per
cent from that used by Marconi \& Hunt.

\item
We also note that Marconi \& Hunt used a distance of 107 Mpc 
for NGC~6251, slightly greater than the value of 101 Mpc
which we adopted in the previous section and which we use here for consistency. 
This results in our reduction of the 
SMBH mass for this galaxy by the fraction 101/107, 
and a dimming of the absolute magnitude by $5\log(107/101)$. 

\end{itemize}

Table~\ref{TabMaH} presents our updated and modified data set from
Marconi \& Hunt's Group 1 galaxies.
Finally, we again note that we perform a non-symmetrical
regression analysis, as is given in equation~\ref{Eq_One}.
Marconi \& Hunt used the symmetrical 
bisector linear regression algorithm of Akritas \& Bershady (1996)
when reporting their optimal relation.   As can be seen in their Figure~1, 
this results in a slightly steeper slope than obtained with an 
ordinary (non-symmetrical) least-squares fit to the data (see also 
Section~\ref{SecDry} of this paper).  Marconi (2007, priv.\ comm.) 
reports that the slope of their data using equation~\ref{Eq_One}
is $-0.41\pm0.04$.

\begin{table}
\caption{
Updated sample of elliptical and disc galaxies from Marconi \& Hunt's (2003)
Group 1 galaxies.  When available, distances from Tonry et al.\ (2001) have
been used. The exceptions are NGC~5252, NGC~6251 and Cygnus~A, 
as noted in Section~\ref{SecMH_K}.
The magnitudes have come from Table~1 in Marconi \& Hunt, adjusted here 
if a different distance was adopted (3 galaxies) or if we assigned a 
bulge-to-disc ratio not used by Marconi \& Hunt (NGC~221 and NGC~4564). 
The source of the black hole masses is also provided in Section~\ref{SecMH_K}. 
\label{TabMaH}
}
\begin{tabular}{@{}lcrccl@{}}
\hline
Galaxy   & Type & Dist.  &  $M_B$  &  $M_K$  &  $M_{\rm bh}$      \\
         &      & [Mpc]  &  [mag]  &  [mag]  & [$10^8 M_{\sun}$]  \\
NGC 0221 &  S   &  0.8   &  -15.3  &  -19.3  &   $0.025^{+0.005}_{-0.005}$ \\
NGC 1023 &  S   & 11.4   &  -18.4  &  -23.5  &   $0.44^{+0.05}_{-0.05}$ \\
NGC 2787 &  S   &  7.5   &  -17.3  &  -21.3  &   $0.41^{+0.04}_{-0.05}$ \\
NGC 3031 &  S   &  3.9   &  -18.2  &  -24.1  &   $0.76^{+0.22}_{-0.11}$ \\
NGC 3115 &  S   &  9.7   &  -20.2  &  -24.4  &   $ 9.2^{+3.0}_{-3.0}$   \\
NGC 3245 &  S   & 20.9   &  -19.6  &  -23.3  &   $2.1^{+0.5}_{-0.5}$  \\
NGC 3377 &  E   & 11.2   &  -19.0  &  -23.6  &   $1.0^{+0.9}_{-0.1}$  \\
NGC 3379 &  E   & 10.6   &  -19.9  &  -24.2  &   $1.35^{+0.73}_{-0.73}$ \\
NGC 3384 &  S   & 11.6   &  -19.0  &  -22.6  &   $0.16^{+0.01}_{-0.02}$ \\
NGC 3608 &  E   & 22.9   &  -19.9  &  -24.1  &   $1.9^{+1.0}_{-0.6}$   \\ 
NGC 4258 &  S   &  7.2   &  -17.2  &  -22.4  &   $0.39^{+0.01}_{-0.01}$ \\
NGC 4261 &  E   & 31.6   &  -21.1  &  -25.6  &   $5.2^{+1.0}_{-1.1}$  \\
NGC 4291 &  E   & 26.2   &  -19.6  &  -23.9  &   $3.1^{+0.8}_{-2.3}$  \\
NGC 4374 &  E   & 18.4   &  -21.4  &  -25.7  &   $4.64^{+3.46}_{-1.83}$ \\
NGC 4473 &  E   & 15.7   &  -19.9  &  -23.8  &   $1.10^{+0.4}_{-0.79}$ \\
NGC 4486 &  E   & 16.1   &  -21.5  &  -25.6  &   $34.3^{+9.7}_{-9.7}$   \\
NGC 4564 &  S   & 15.0   &  -17.4  &  -21.9  &   $0.56^{+0.03}_{-0.08}$  \\
NGC 4594 &  S   &  9.8   &  -21.3  &  -25.4  &   $10.0^{+10.0}_{-7.0}$  \\
NGC 4649 &  E   & 16.8   &  -21.3  &  -25.8  &   $20.0^{+4.0}_{-6.0}$   \\
NGC 4697 &  E   & 11.7   &  -20.2  &  -24.6  &   $1.7^{+0.2}_{-0.1}$ \\
NGC 4742 &  E   & 15.5   &  -18.9  &  -23.0  &   $0.14^{+0.04}_{-0.05}$ \\
NGC 5128 &  S   &  4.2   &  -20.8  &  -24.5  &   $2.4^{+3.6}_{-1.7}$  \\
NGC 5252 &  S   & 94.4   &  -20.7  &  -25.5  &   $ 9.7^{+14.9}_{-4.6}$ \\
NGC 5845 &  E   & 25.9   &  -18.7  &  -23.0  &   $2.4^{+0.4}_{-1.4}$ \\
NGC 6251 &  E   & 101    &  -21.4  &  -26.5  &   $5.8^{+1.8}_{-2.0}$  \\
Cygnus A &  E   & 240    &  -21.8  &  -27.2  &   $26.0^{+7.0}_{-7.0}$  \\
\hline
\end{tabular}
\end{table}

\begin{figure}
\includegraphics[angle=270,scale=0.64]{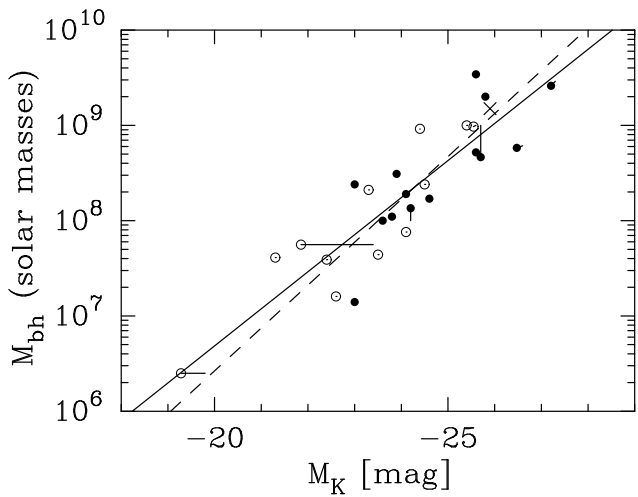}
\caption{The long solid line shows the $M_{\rm bh}$-$L$ 
relation (equation~\ref{Eq_MH_K}) using our updated values
for the galaxies in Marconi \& Hunt (2003). 
The short lines emanating from the data points show the location 
of the galaxies as used by Marconi \& Hunt to obtain 
the long dashed line.  The location of IC~1459 as used by Marconi \&
Hunt but excluded by us is shown by the cross. 
}
\label{FigMH}
\end{figure}

Figure~\ref{FigMH} shows the $K$-band $M_{\rm bh}$-$L$ relation
derived using our slightly updated Marconi \& Hunt data set. 
One can also see, via the short lines, 
how each data point has moved from its previous location as given 
by Marconi \& Hunt.  Performing a linear regression (using 
equation~\ref{Eq_One}), we obtain
\begin{equation}
\log(M_{\rm bh}/M_{\sun}) = -0.39(\pm0.05)[M_K + 24] + 8.24(\pm0.08), 
\label{Eq_MH_K}
\end{equation}
with a total scatter of 0.35 dex in $\log M_{\rm bh}$. 
This equation has a slightly shallower slope
than the expression given by Marconi \& Hunt (2003, their Table~2): 
$\log(M_{\rm bh}/M_{\sun}) = 1.13(\pm0.12)[\log(L_{K,{\rm bulge}}/L_{K,\sun}) -10.9] + 8.21(\pm0.07)$,
or equivalently\footnote{Dong \& De Robertis (2006) obtain the same
slope ($-0.45$) upon excluding the disc galaxies, but with a large
uncertainty on the intercept.}, using $M_{K,\sun}=3.28$ mag, 
$\log(M_{\rm bh}/M_{\sun}) = -0.45(\pm0.05)[M_{K,{\rm bulge}} + 23.97] +
8.21(\pm0.07)$.   However, as noted above, 
applying the same regression analysis to the 
data in Marconi \& Hunt yields a consistent result.

\subsubsection{Removing questionable data points} \label{Sec22}

We repeated the above regression analysis removing four galaxies whose
parameters are somewhat questionable.  
First, because the mass estimate for NGC~4742 
has not yet appeared in a refereed paper, we hold off on its inclusion 
here.  We also excluded NGC~1023 ($r_h=0^{\prime\prime}.08$) and 
NGC~3377 ($r_h=0^{\prime\prime}.46$) because their 
SMBH spheres of influence ($r_h$, Merritt \& Ferrarese 2001c) 
have apparently not been resolved according to Table~II in Ferrarese \& Ford (2005) 
in which $r_h/r_{res}$ ratios of 0.89 and 0.74 respectively. 
Finally, as noted by Ferrarese \& Ford (2005), NGC~4594 has not yet had
its SMBH mass acquired using a 3-integral model, and so it mass 
may therefore be in error.  
With our reduced sample of 22 objects, we obtain the relation 
\begin{equation}
\log(M_{\rm bh}/M_{\sun}) = -0.37(\pm0.04)[M_K + 24] + 8.29(\pm0.08), 
\label{Eq_MH_Ali}
\end{equation} 
consistent with the expression given in equation~\ref{Eq_MH_K}. 
The total scatter about this relation is 0.33 dex, and the intrinsic 
scatter is $0.30^{+0.03}_{-0.05}$ dex in $\log M_{\rm bh}$. 

Comparison of equation~\ref{Eq_MH_Ali} (and \ref{Eq_MH_K}) 
with our revised estimate of McLure \& Dunlop's $K$-band 
relation (equation~\ref{Eq_MD17K}) 
reveals that we have resolved the 
disagreement noted in our Introduction. 
That is, our updated data sets and reanalysis of the 
McLure \& Dunlop and the Marconi \& Hunt studies has yielded
$M_{\rm bh}$-$L$ relations that agree with each other. 
Our preference is to use equation~\ref{Eq_MH_Ali} because 
a) it was derived using both elliptical and disc galaxies and
b) it has the smallest uncertainty on the slope and intercept.

\subsubsection{Is the $M_{\rm bh}$-$L$ relation non-linear} \label{SecQuad}

Using the above 22 data points, we have explored whether the 
$M_{\rm bh}$-$L$ relation may be curved.  The optimal 
log-quadratic relation, fitted in the same way as the $M_{\rm bh}$-$n$ data 
in Graham \& Driver (2007a), is 
\begin{eqnarray}
\log(M_{\rm bh}/M_{\sun}) & = &
-0.37(\pm0.05)[M_K + 24] + 8.32(\pm0.10) \nonumber \\
 & & -0.01(\pm0.02)[M_K + 24]^2. 
\label{Eq_Quad}
\end{eqnarray}
The coefficient in front of the quadratic term is consistent with a 
value of zero, or in other words, it does not deviate from a value
of zero at even the 1-sigma level.  The 3$\sigma$ range of values on 
this term is only $\pm0.05$.  One can therefore conclude that the 
$M_{\rm bh}$-$L$ relation, defined with the present data set, is not curved.

\subsubsection{Is the $M_{\rm bh}$-$L$ relation the same for elliptical galaxies and bulges} \label{SecES}

For the 12 elliptical galaxies which comprise the sample of 22 galaxies
used in Section~\ref{Sec22}, the best-fitting relation is
\begin{equation}
\log(M_{\rm bh}/M_{\sun}) = -0.33(\pm0.09)[M_K + 24] + 8.33(\pm0.15). 
\label{Eq_MH_E}
\end{equation}
For the 10 disc galaxies, which includes NGC~221 and NGC~4564, one has
\begin{equation}
\log(M_{\rm bh}/M_{\sun}) = -0.39(\pm0.08)[M_K + 24] + 8.33(\pm0.16). 
\label{Eq_MH_S}
\end{equation}
Consequently, there appears to be no significant difference between the
relations defined by the elliptical galaxies and the bulges of disc galaxies.

\subsubsection{An $R$-band $M_{\rm bh}$-$L$ relation} \label{SecMH_R}

The above sample of 22 galaxies consists of 12 E galaxies,
8 S0 galaxies, and only one Sb and one Sbc galaxy.
Using an $R_c-K$ colour of 2.6 and 2.5 for the
elliptical and lenticular galaxies respectively, 
and 2.3 for the late-type galaxies (Buzzoni 2005)
we obtained the following $R$-band $M_{\rm bh}$-$L$ relation 
\begin{equation}
\log(M_{\rm bh}/M_{\sun}) = -0.38(\pm0.04)[M_R + 21] + 8.12(\pm0.08),
\label{Eq_MH_R}
\end{equation}
While this relation has overlapping error bars with the two independent 
$R$-band relations 
from the previous sections (equation~\ref{Eq_MD17R} and \ref{Eq_Ali}), 
it is preferred for the reasons mentioned in Section~\ref{Sec22}.

\subsubsection{A $B$-band $M_{\rm bh}$-$L$ relation}

Starting with the $B$-band magnitudes tabulated in Marconi \& Hunt, 
(which have predominantly come from Tremaine et al.\ 2002 via
Faber et al.\ 1997), we have modified these according
to Section~\ref{SecMH_K}.  Adjusting also the SMBH masses as in
Section~\ref{SecMH_K}, we obtain the updated $B$-band expression 
\begin{equation}
\log(M_{\rm bh}/M_{\sun}) = -0.40(\pm0.05)[M_B + 19.5] + 8.27(\pm0.08), 
\label{Eq_MH_B}
\end{equation}
with a total scatter of 0.34 dex in $\log M_{\rm bh}$. 
Using $M_{B,\sun}=5.47$ mag (Cox 2000), equation~\ref{Eq_MH_B} 
has a marginally shallower slope than the solution in Marconi \& Hunt: 
$\log(M_{\rm bh}/M_{\sun}) = -0.48(\pm0.05)[M_B + 19.53] + 8.18(\pm0.08)$. 

This new $B$-band slope of $-0.40\pm0.05$ (obtained using 22 galaxies) is 
comparable to the slope $-0.42\pm0.06$ (equation~\ref{Eq_MD_M32}) 
obtained in Section~\ref{SecMD} 
using M32 plus the 16 elliptical galaxies from McLure \& Dunlop (2002).
However, given no correction for dust attenuation was used, and given that the
past $B$-band bulge/disc separation could probably be improved upon, this may
perhaps be fortuitous.

\subsubsection{Scatter in the $M_{\rm bh}$-$M_{\rm spheroid}$ relation}

Marconi \& Hunt (2003) additionally presented a relation between the
mass of the black hole and the mass of the host spheroid.  They used
$M_{\rm spheroid} = 3 R_{\rm e} \sigma_{\rm e}^2/G$, in which $R_{\rm e}$
and $\sigma_{\rm e}$ are the effective half-light radius and velocity 
dispersion of each spheroid, respectively. 
They reported (for their 27 Group 1 galaxies) an intrinsic scatter of 0.25 dex 
(and 0.49 dex for the full sample).   Using their Group 1 data, and fitting
equation~\ref{Eq_One} --- which is designed to minimise the scatter in $\log M_{\rm
  bh}$ --- we measure the {\it total} scatter to be 0.30 dex.
A fuller, proper investigation
of this subject would however require checking the $R_{\rm e}$ 
values, which is beyond the scope of the current paper.



\section{Discussion and Conclusions}

Surprisingly, given the recent large body of work on SMBHs, there remains a
strong need for high quality images for the sample
of inactive galaxies with direct measurements of their supermassive black hole
masses.  In particular, near-infrared \textit{NICMOS} images would enable an
analysis of the core structure of these galaxies.  Graham \& Driver (2007a)
have proposed that the central stellar density may be the fundamental parameter
related to the mass of the black hole.  
This contrasts with current studies, including this one, 
which explore the SMBH connection with global rather 
than nuclear properties of the host spheroid. Given the known trend between
central stellar density and host 
spheroid luminosity (e.g., Graham \& Guzm\'an 2003; Merritt 2006, his Fig.5), 
the popular relations {\it may} all be secondary in nature. i.e., subsequential
Specifically, the central density
  (prior to core-depletion\footnote{The apparent depletion of stars at the
  centres of giant galaxies may not have (only) arisen from the scouring
  action of coalescing SMBHs (see, for example, Boylan-Kolchin et al. 2004 and
  Nipoti et al.\ 2006).}  in massive spheroids  --- requiring the use of the
core-S\'ersic model, Graham et al.\ 2003), or the central density of less
massive spheroids after modelling and excluding the flux from their additional nuclear
components (e.g. Graham \& Guzm\'an 2003) may be
the key parameter connected to the SMBH mass.  Due to the need for a bulge/disc
decomposition for half of the current galaxy sample, such \textit{HST} images
should be mated with deep larger field-of-view ground-based images (e.g.
Balcells et al.\ 2004) so as to adequately sample the domain of the disc.

Aside from the study by Erwin et al.\ (2004) with 13 galaxies, all of the
optical $M_{\rm bh}$-$L$ relations to date have been constructed using the
aperture growth curve magnitudes obtained some 20 years ago and presented in
Faber et al.\ (1989) or from the RC3 (de Vaucouleurs et al.\ 1991).
%
A homogeneous set of high-resolution, deep, wide-field CCD images in 
the optical bands such as $B$ and $R$ would be highly useful for
a) properly calibrating the $M_{\rm bh}$-$L$ relation, b) acquiring
accurate bulge sizes and central densities (both projected and deprojected)
and c) subsequently calibrating the $M_{\rm bh}$-(spheroidal mass) relation
(e.g.\ Marconi \& Hunt 2003; H\"aring \& Rix 2004).  Not 
only would this allow a proper test for the optimal fundamental relation, 
but it would provide the community with improved relations for predicting 
SMBH masses in other galaxies. 

Nonetheless, there does now appear to be agreement between the various
$M_{\rm bh}$-$L$ relations (see Table~\ref{TabFin}).

\begin{table*}
 \centering
 \begin{minipage}{125mm}
\caption{
New $M_{\rm bh}$-$L$ relations for predicting SMBH masses. 
The ``origin'' of the data is as follows.
KG 2001 = Kormendy \& Gebhardt (2001), not modified.
MD 2002 = McLure \& Dunlop (2002), modified (see Section~\ref{SecMD} and 
Table~\ref{TabMD}).
EGC 2004 = Erwin et al.\ (2004), modified (see Section~\ref{SecEGC} and 
Table~\ref{Tab_EGC}).
MH 2003 = Marconi \& Hunt (2003), modified (see Section~\ref{SecEGC}).
and Table~\ref{TabMaH}). 
The ``sample'' may consist of elliptical (E) galaxies or disc galaxies,
denoted by `S' for either S0 or Sp.
The total scatter in the $\log M_{\rm bh}$ direction is denoted 
$\Delta_{\rm tot}$, while the intrinsic scatter in the 
$\log M_{\rm bh}$ direction is denoted $\epsilon_{\rm intrinsic}$. 
\label{TabFin}
}
\begin{tabular}{@{}llcccc@{}}
\hline
Origin & Sample & Band & $M_{\rm bh}$-$L$ relation & $\Delta_{\rm tot}$ & $\epsilon_{\rm intrinsic}$ \\
       &        &      &                           &  [dex]             & [dex]  \\
KG 2001 & 20E + 17S & $B$ & $-0.38(\pm0.06)[M_B+19.5] +8.00(\pm0.09)$ & 0.56 & $0.46^{+0.08}_{-0.06}$ \\
MD 2002 & 16E + 1S  & $B$ & $-0.36(\pm0.06)[M_B+20] +8.33(\pm0.10)$ & 0.38   & $0.35^{+0.03}_{-0.06}$ \\
MH 2003 & 12E + 10S & $B$ & $-0.40(\pm0.05)[M_B+19.5] +8.27(\pm0.08)$ & 0.34 & $0.30^{+0.04}_{-0.05}$ \\
MD 2002 & 16E + 1S  & $R$ & $-0.38(\pm0.06)[M_R+21] +8.11(\pm0.11)$ & 0.38 & $0.35^{+0.03}_{-0.07}$ \\
EGC 2004 & 08E + 5S & $R$ & $-0.30(\pm0.06)[M_R+21] +7.96(\pm0.10)$ & 0.31 & $0.28^{+0.03}_{-0.06}$ \\
MH 2003 & 12E + 10S & $R$ & $-0.38(\pm0.04)[M_R+21] +8.12(\pm0.08)$ & 0.33 & $0.30^{+0.03}_{-0.05}$ \\
MD 2002 & 16E + 1S  & $K$ & $-0.38(\pm0.06)[M_K+24] +8.26(\pm0.11)$ & 0.38 & $0.35^{+0.03}_{-0.07}$ \\
MH 2003 & 12E + 10S & $K$ & $-0.37(\pm0.04)[M_K+24] +8.29(\pm0.08)$ & 0.33 & $0.30^{+0.03}_{-0.05}$ \\
\hline
\end{tabular}
\end{minipage}
\end{table*}

We plan to apply our $B$-band $M_{\rm bh}$-$L$ relation
(equation~\ref{Eq_MH_B}) to the Millennium Galaxy Catalogue (e.g., Driver et al.\ 
2006).  This catalogue contains structural parameters from the $R^{1/n}$-bulge
plus exponential-disc decomposition of 10,095 nearby ($z\sim0.1$) galaxies
(Allen et al.\ 2006).  While we have already been able to use the $M_{\rm
  bh}$-$n$ relation in Graham \& Driver (2007a) to predict the SMBH masses in 
these galaxies (Graham et al.\ 2007), the inconsistencies in the previously 
published $M_{\rm bh}$-$L$ relations had prohibited their use.  
This will allow us to 
construct an updated SMBH mass function for both early- and late-type 
galaxies, which can then be integrated to obtain the 
local SMBH mass density.

\subsection{Intermediate mass black holes}

\begin{figure}
\includegraphics[angle=270,scale=0.64]{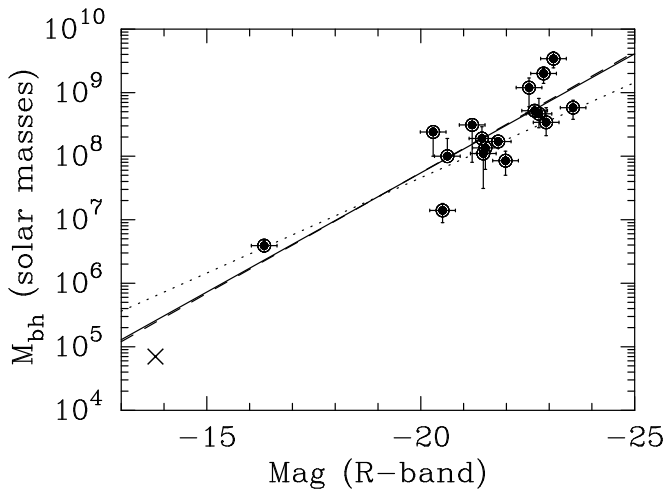}
\caption{
SDSS J160531.84+174826.1 (cross) has been 
added to the data points used to construct equation~\ref{Eq_MD17R}, shown here 
by the solid line. This dwarf galaxy, from Dong et al.\ (2007), has an intermediate 
mass black hole that has not been used in the fitting of the lines shown. 
The dashed line is equation~\ref{Eq_MH_R} and the dotted line is
equation~\ref{Eq_Ali}. 
}
\label{Fig_Dong}
\end{figure}

Using the linewidth-luminosity-mass scaling relation by Greene \& Ho
(2005), Dong et al.\ (2007) report on the existence of a $7\times 10^4 M_{\sun}$ 
intermediate mass black hole (IMBH) in the dwarf disk galaxy SDSS 
J160531.84+174826.1 (see Figure~\ref{Fig_Dong}). 
Applying McLure \& Dunlop's (2002) $R$-band $M_{\rm bh}$-$L$ relation 
to the central bulge/bar magnitude of this disk galaxy, 
they obtained a black hole mass one order of magnitude 
smaller than obtained with Marconi \& Hunt's (2003) relation. 
Dong et al.\ give a ($H_0=70$ km s$^{-1}$ Mpc$^{-1}$) 
$R$-band magnitude of $-13.9$ mag. 
Adjusting this by 0.1 mag, to match our
adopted value of $H_0 =73$ km s$^{-1}$ Mpc$^{-1}$, 
our updated $R$-band expression from equation~\ref{Eq_MD17R} predicts 
a black hole mass of $2.4\times10^5 M_{\sun}$, while our
updated relation in equation~\ref{Eq_MH_R} also gives a value of 
$2.4\times10^5 M_{\sun}$.  This compares well 
with the value of $1.5\times10^5 M_{\sun}$ obtained using the 
$M_{\rm bh}$-$n$ relation from Graham \& Driver (2007a). 


%
We do not include the Seyfert 1 galaxy POX~52 as it appears to require 
a (yet to be performed) bulge/disc decomposition, with the disk accounting
for the excess flux from 4-10 arcseconds seen in the S\'ersic fit of Barth 
et al.\ (2004). 
%
%
%
Ideally, a bulge/disc decomposition is also required for the dwarf Seyfert 1,
Sd galaxy NGC~4395.  
Inspection of figure~3 in Filippenko \& Ho (2003) suggests that the 
excess flux seen from 1 to $\sim$2 arcseconds, peaking at $\sim$0.6 
mag arcsec$^{-2}$ above their fitted model, may be the bulge component
of this late-typpe galaxy.  However, 
due to the non-thermal point-source emission and nuclear-star cluster
in this galaxy, it is difficult to know what may be the bulge component. 
While it appears that a Gaussian-like component should work well for the 
suggested 
bulge, we make no attempt to undertake this task here, but highlight the value of
such a future investigation.   The addition of more data points at the 
low-mass end of the $M_{\rm bh}$-$L$ relation should provide further valuable
clues as to a) the range and reliability of this relation, and b) help
constrain models for the co-evolution of black holes and galaxies.

\subsection{Implications for SMBH-galaxy coevolution}\label{SecDry}

Before discussing intrinsic physical relations, we need to 
perform a symmetrical regression of $M_{\rm bh}$ and the magnitude.
We have already seen that regressing $M_{\rm bh}$ on $M_K$
gives a $K$-band slope of $-0.37\pm0.04$ (equation~\ref{Eq_MH_Ali}).
Reversing the regression (see the text after equation~\ref{Eq_One}) gives the
relation $\log(M_{\rm bh}/M_{\sun}) = -0.44(\pm0.05)[M_K + 24] +
8.30(\pm0.09)$, and so the average slope is $\sim 0.40$.
This is consistent with the slope of $-0.45\pm0.05$ from the
symmetrical regression in Marconi \& Hunt (2003). 
It is interesting to note that a slope of $-0.40$ implies 
a linear scaling between luminosity and black hole mass, such that 
$M_{\rm bh} \propto L^{1.00}$.  That is, the SMBH mass to 
spheroid luminosity ratio is constant, 
as predicted by Chien (2007). 

Our updated data and new relation supports the ``dry merger'' 
scenario in which the SMBH mass doubles as the luminosity doubles. 
While this paints a consistent picture at the high-mass end,
where ``core galaxies'' exist and dry mergers are considered the 
order of the day, this is unlikely to hold at lower masses. 

SMBH accretion of ISM gas derived from stellar winds is believed to fuel some
AGN activity today (Fabbiano et al.\ 2004, Pellegrini 2005; 
Soria et al.\ 2006).  Indeed, stellar mass loss in
elliptical galaxies produces $>10^2$ times more mass than that found in the
central SMBH (Ciotti \& Ostriker 2007).  As these Authors note, when fuelling
SMBHs with the recycled gas from the stellar population, almost by definition,
the amount of fuel is proportional to the host galaxy's stellar mass.  Ciotti
\& Ostriker also show how radiative heating from such AGN feedback is
responsible for the self-regulated coevolution of galaxy and SMBH.  
%
%
Our results provide a valuable new constraint on these models, setting the 
proportionality constant to one.


Moreover, the level of total scatter in the $M_{\rm bh}$-$L$ relation (0.31 to
0.34 dex) makes it competitive with both 
the $M_{\rm bh}$-$n$ (Graham et al.\ 2001; Graham \& Driver 2007a: 0.31 dex) and 
the $M_{\rm bh}$-$\sigma$ relation (Ferrarese \& Merritt 2000; Gebhardt et
al.\ 2000).  While modelling the 31 galaxies from Tremaine et al.\ (2002) 
gives an intrinsic scatter of 0.27 dex, the total scatter about the
relation presented there is 0.34 dex.

This opens the question as to whether the stellar (baryonic) 
or total (stars plus dark matter) mass is the driver of the SMBH-galaxy 
connection.  Which of these, or some other quantity may be the 
fundamental parameter connecting galaxies and their black holes is not yet
clear.

%
%
%

\section{acknowledgments}

The Author is grateful to Simon Driver for useful discussions while investigating the 
McLure \& Dunlop data set, and to Ana Matkovi\'c for providing the Coma
data used in Figure~\ref{Fig_Lsigma}.  The Author is also happy to acknowledge 
useful discussions with Alessandro Marconi, Ross McLure and Mariangela Bernardi. 
This research has made use of the NASA/IPAC Extragalactic Database (NED).

\appendix

\section{Potential sample bias} \label{SecBias}


Bernardi et al.\ (2007) have suggested that the local sample of inactive
galaxies with direct SMBH mass measurements may be biased with regard to the
greater population.  They argue that, relative to the total population, 
there is a bias such that these 
galaxies have overly large velocity dispersions for their luminosities. 
If correct, the result is such that the
$M_{\rm bh}$--$L$ relation will over-predict the SMBH masses (in other
galaxies) relative to the $M_{\rm bh}$--$\sigma$ relation.
We explore this issue with our updated data set. 

\begin{figure}
\includegraphics[angle=270,scale=0.48]{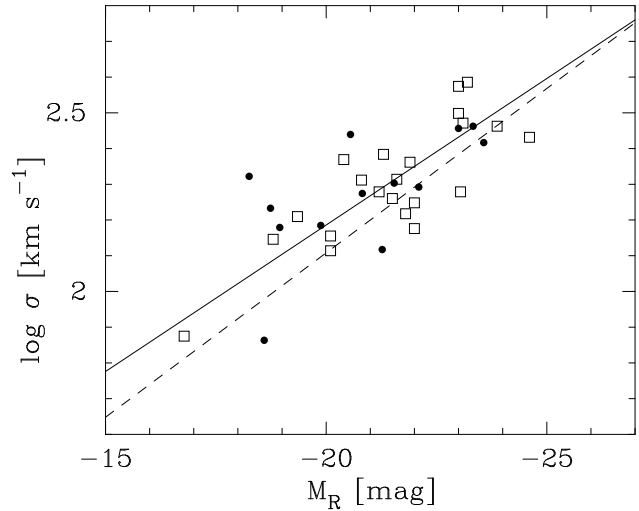}
\caption{
The solid line (equation~\ref{Eq_MH_LSig}) shows the regression of $\log \sigma$ on $M_R$
using our updated Marconi \& Hunt data set, denoted by the open squares.
Our updated Erwin et al.\ data points are shown by the filled circles
but are not used in the linear regression shown here.  The dashed line 
(equation~\ref{Eq_SDSS}) is the equivalent regression using SDSS data 
(Tundo et al.\ 2007, their equation~4). 
A slight offset is evident. 
}
\label{Fig_sigmaL}
\end{figure}

Using $r-R=0.24$ (Fukugita 1995), the SDSS $L$-$\sigma$ relation from Tundo 
et al.\ (2007, their equation~4) is such that
\begin{equation}
\log \sigma = 0.27 - 0.092M_R. 
\label{Eq_SDSS}
\end{equation}
We compare this with our updated 
$R$-band data from Erwin et al.\ (2004) and 
our updated $K$-band data from Marconi \& Hunt transformed 
into the $R$-band using the tables in Buzzoni (2005), as done in
Section~\ref{SecMH_R}. 
The results are shown in Figure~\ref{Fig_sigmaL}. 

Applying equation~\ref{Eq_One} to our updated Marconi \& Hunt data set,
in which we have minimised the scatter in the $\log \sigma$ direction, 
we obtain 
\begin{equation}
\log \sigma = 2.268 - 0.082(M_R +21), 
\label{Eq_MH_LSig}
\end{equation}
which is shown by the solid line in Figure~\ref{Fig_sigmaL}. 
Compared to the dashed line (from the SDSS data set), 
the local sample of inactive galaxies therefore appears to have 
larger velocity dispersions for a given magnitude.  

While the local sample of galaxies with direct SMBH mass measurements
may be biased, it is also possible that the magnitudes used to construct the 
SDSS $L$-$\sigma$ relation may have been under-estimated due to dust, 
or over-estimated due to the bulge-disc separation. 
That is, {\it they} may be the biased data set. 
Unlike the Erwin et al.\ (2004) and Marconi \& Hunt (2003) bulge 
data, no S\'ersic $R^{1/n}$-bulge $+$ exponential-disc fit was 
performed in acquiring the SDSS bulge magnitudes.  The use of $R^{1/4}$ 
models is known to over-estimate the bulge flux for bulges with
stellar distributions having $n<4$ (e.g. Graham \& Driver 2007b, 
their Table~3), which may be the bulk of the S0 galaxy 
population (e.g., Balcells et al.\ 2003). 
A second concern is that faint elliptical galaxies have a
different slope in the $L$-$\sigma$ diagram than luminous
elliptical galaxies.  The successive inclusion of fainter 
galaxies should therefore lead to a progressive change in 
the slope of the $L$-$\sigma$ relation.

Ignoring the above issues for now, the average velocity dispersion 
excess is given by differencing the above two equations, to give
$\Delta \log \sigma = 0.30 + 0.011M_R$. 
Therefore, if one wished to correct for this (possible) sample bias, 
the local galaxy magnitudes $M_R$ (i.e., those with direct
SMBH mass measurements) could be adjusted by 
$\Delta M$, such that 
\begin{equation}
\Delta M = 0.109[M_R+27.60]. 
\label{Eq_Dhigh}
\end{equation}
This would bring the two lines in Figure~\ref{Fig_sigmaL} into agreement.
The same trick was performed in Tundo et al.\ (2007, at the end of their
Section~3), although using different bulge magnitudes to our updated values. 
In a future paper we intend to better address such a potential offset in the
$L$--$\sigma$ diagram using the dust-corrected $R^{1/n}$-bulge magnitudes from
the Millennium Galaxy Catalogue's 10 095 galaxies (Allen et al.\ 2006).

At the low luminosity end the situation is different to 
presented above.  This is because the $L$--$\sigma$ 
relation is not linear, having a well-recognised slope of $\sim$4 at the
bright end and a less well known, but long established, slope of $\sim$2 at
the faint end (Tonry 1981; Davies et al.\ 1983; Held et al.\ 1992; De Rijcke
et al.\ 2005).  Matkovi\'c \& Guzm\'an (2005) argue that the transitional
magnitude occurs near $M_R \sim -22$ mag, coinciding with the onset of dry
merging and the break seen in the luminosity - central surface brightness
diagram (Graham \& Guzm\'an 2003, their figure~9c).  What is also apparent in
Figure~4 from Matkovi\'c \& Guzm\'an is the subtle nature of this transition,
and the need for a long baseline in magnitude for it to be recognised.

Fitting a {\it single} power-law to the $L$-$\sigma$ data should yield a slope that
is dependent on one's luminosity range and thus sample selection.  This has
not received much attention in the literature, most likely because of the
subtle nature of the transition due to the scatter about the relation.  For
this reason, the change in slope only really becomes obvious in samples
containing magnitudes fainter than $M_R \sim -20$ mag.

\begin{figure}
\includegraphics[angle=270,scale=0.48]{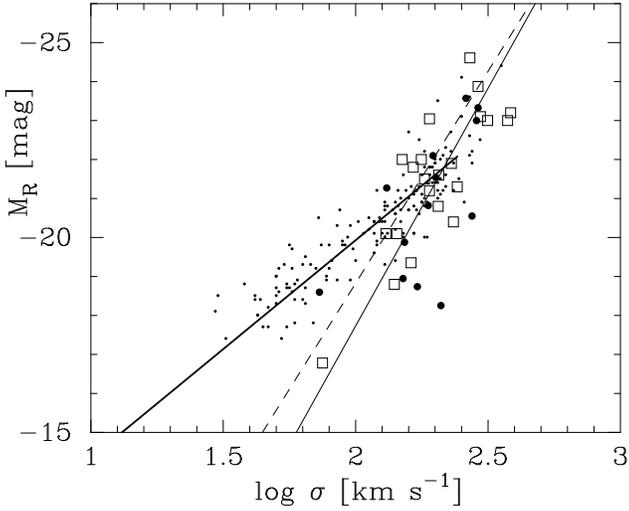}
\caption{
The solid and dashed lines have the same meaning as in
Figure~\ref{Fig_sigmaL}, as do the squares and large filled circles.
The small dots are data taken from Matkovi\'c \& Guzm\'an (2005, their
Figure~4, adjusted from $H_0$=70 to 73 km s$^{-1}$ Mpc$^{-1}$)  
and the thick solid line is their $H_0$-adjusted ($\log \sigma \mid M_R$) 
regression for faint galaxies. 
}
\label{Fig_Lsigma}
\end{figure}

Performing a regression of $\log \sigma$ on $M_R$, Matkovi\'c \& 
Guzm\'an (2005) report a faint-end relation of
\begin{equation}
M_R = (-5.585\pm0.210)\log \sigma - 8.755(\pm0.444), \nonumber
\end{equation}
or simply
\begin{equation}
\log \sigma = -0.179M_R - 1.508, 
\label{Eq_MG05}
\end{equation}
which has been adjusted here to $H_0$=73 km s$^{-1}$ Mpc$^{-1}$. 
Their data and this relation can be seen in Figure~\ref{Fig_Lsigma},
along with our sample of galaxies with direct SMBH mass measurements.
One can see that the low-luminosity extrapolation of the $L$-sigma relation 
obtained using the updated Marconi \& Hunt data set 
(equation~\ref{Eq_MH_LSig}) does not follow the trend defined 
by the fainter galaxy population (equation~\ref{Eq_MG05}). 
The problem is such that for a given velocity dispersion, the 
magnitudes predicted from the (SMBH sample)-derived $L$-$\sigma$ relation
are too faint, compared to the general population. 
The magnitudes predicted using equation~\ref{Eq_MH_LSig} need to brightened
by 
\begin{equation}
\Delta M = 0.542[M_R+21.78]. 
\label{Eq_Dlow}
\end{equation}

If the sample of galaxies with direct SMBH masses
have normal masses and velocity dispersions, but biased luminosities, 
as argued by Bernardi et al.\ (2007), then 
one can apply the above magnitude corrections to this sample. 
At $M_R = -20.28$ mag we switch from using corrective equation~\ref{Eq_Dhigh} to 
corrective equation~\ref{Eq_Dlow}.   With our modified set of galaxy magnitudes, 
corrected to represent the general population, one can repeat the 
regression analysis to obtain a new $M_{\rm bh}$-$L$ relation.
Doing so, one has 
\begin{equation}
\log(M_{\rm bh}/M_{\sun}) = -0.51(\pm0.06)[M_R + 22] + 8.16(\pm0.09),
\label{Eq_Bias}
\end{equation}
with a total scatter of 0.36 dex, 
and an intrinsic scatter of 0.31$^{+0.05}_{-0.06}$. 

\begin{figure}
\includegraphics[angle=270,scale=0.63]{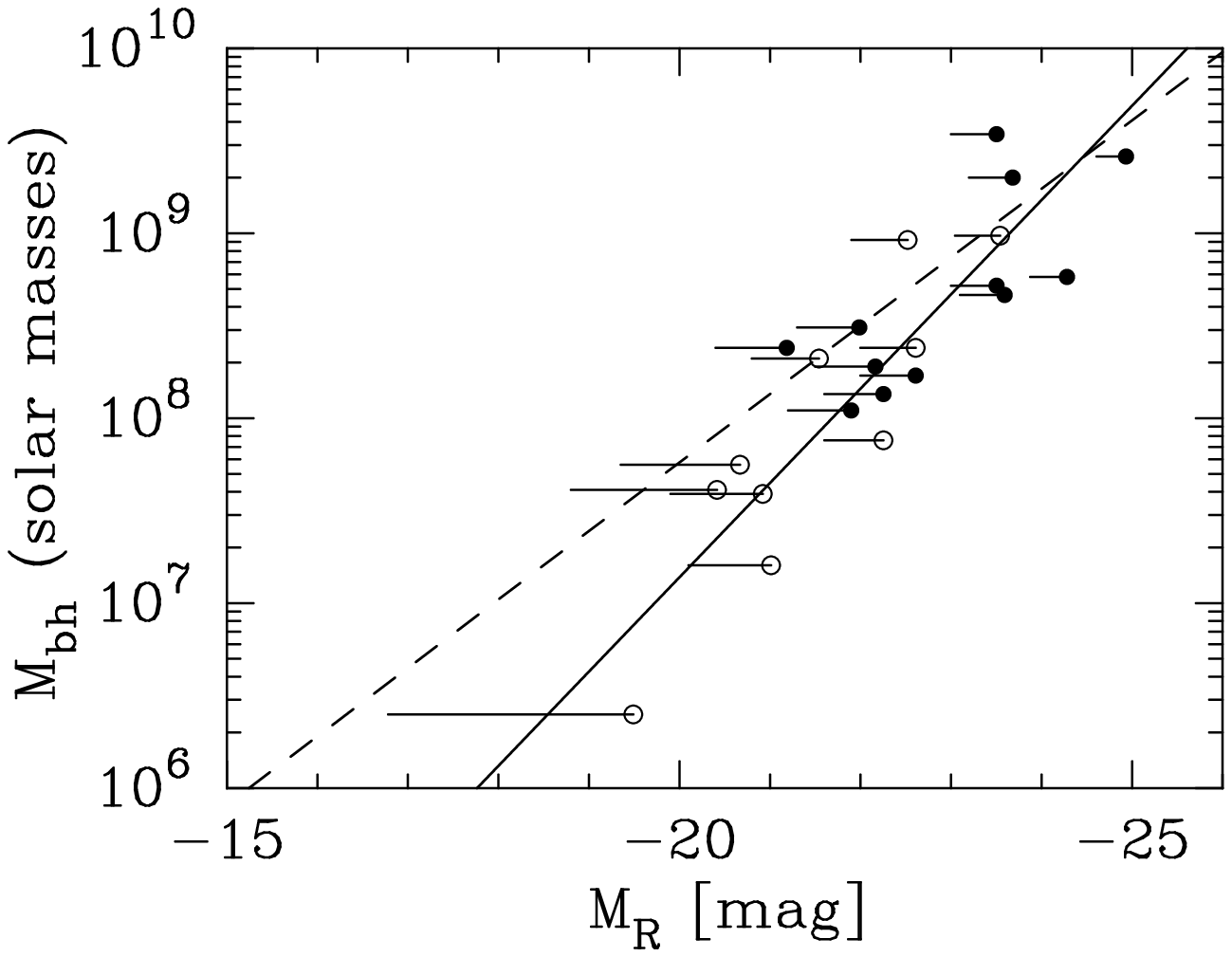}
\caption{
The data points have the same meaning as in Figure~\ref{FigMH}, although 
here we only show those 22 galaxies from Section~\ref{Sec22}.
The short lines emanating from each data point show their location
prior to the magnitude-adjustment performed in Section~\ref{SecBias}.
The location of the data points show their magnitude-adjusted location. 
The dashed line shows the regression prior to this adjustment 
(equation~\ref{Eq_MH_R}) while the solid line shows the new regression
(equation~\ref{Eq_Bias}). 
}
\label{Fig_adjust}
\end{figure}

Compared to equation~\ref{Eq_MH_R}, which was obtained prior to this magnitude
adjustment, for magnitudes fainter than $M_R \sim -24.5$ mag 
the new relation predicts smaller SMBH masses (see Figure~\ref{Fig_adjust}).  
Extrapolation of the 
relations to brighter magnitudes results in greater SMBH masses.
This is a consequence of the $L$-$\sigma$ relations from Figure~\ref{Fig_sigmaL}
that we have used to derive the magnitude adjustment in equation~\ref{Eq_Dhigh}).

An alternative scenario is that the $M_{\rm bh}$-$L$ relation may 
have two slopes, described by a broken power-law with the 
transition denoting the onset of dry merging.  Indeed, the above 
prescription should generate such a relation. 
If one accepts that $M_{\rm bh} \propto \sigma^4$,
then at the luminous end, where $L \propto \sigma^4$, one naturally
obtains $M_{\rm bh} \propto L^{1.0}$.  For magnitudes fainter than 
$M_R \sim -22$ mag, one has $L \propto \sigma^2$, and if $M_{\rm bh} \propto \sigma^4$
over this domain, then one should expect to find
$M_{\rm bh} \propto L^{0.5}$.  However, we feel that to properly 
address this scenario will require more data than is available at present.

\label{lastpage}
\end{document}